\documentclass[8.5pt,twoside,twocolumn]{article}
\usepackage[super,sort&compress,comma]{natbib} 
\usepackage{mhchem}
\usepackage{times,mathptm}
\usepackage{sectsty}
\usepackage{balance} 

\usepackage{graphicx} %eps figures can be used instead
\usepackage{lastpage}
\usepackage[format=plain,justification=raggedright,singlelinecheck=false,font=small,labelfont=bf,labelsep=space]{caption} 
\usepackage{fancyhdr}
\usepackage{color}

\begin{document}

\thispagestyle{plain}
\fancypagestyle{plain}{
\fancyhead[L]{\includegraphics[height=8pt]{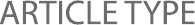}}
\fancyhead[C]{\hspace{-1cm}\includegraphics[height=20pt]{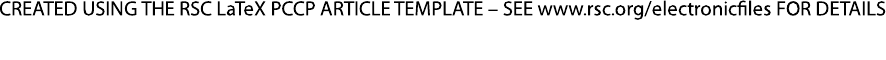}}
\fancyhead[R]{\includegraphics[height=10pt]{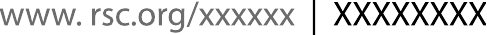}\vspace{-0.2cm}}
\renewcommand{\headrulewidth}{1pt}}
\renewcommand{\thefootnote}{\fnsymbol{footnote}}
\renewcommand\footnoterule{\vspace*{1pt}% 
\hrule width 3.4in height 0.4pt \vspace*{5pt}} 
\setcounter{secnumdepth}{5}

\makeatletter 
\def\subsubsection{\@startsection{subsubsection}{3}{10pt}{-1.25ex plus -1ex minus -.1ex}{0ex plus 0ex}{\normalsize\bf}} 
\def\paragraph{\@startsection{paragraph}{4}{10pt}{-1.25ex plus -1ex minus -.1ex}{0ex plus 0ex}{\normalsize\textit}} 
\renewcommand\@biblabel[1]{#1}            
\renewcommand\@makefntext[1]% 
{\noindent\makebox[0pt][r]{\@thefnmark\,}#1}
\makeatother 
\renewcommand{\figurename}{\small{Fig.}~}
\sectionfont{\large}
\subsectionfont{\normalsize} 

\fancyfoot{}
\fancyfoot[LO,RE]{\vspace{-7pt}\includegraphics[height=9pt]{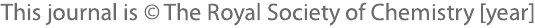}}
\fancyfoot[CO]{\vspace{-7.2pt}\hspace{12.2cm}\includegraphics{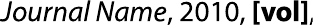}}
\fancyfoot[CE]{\vspace{-7.5pt}\hspace{-13.5cm}\includegraphics{headers/RF.pdf}}
\fancyfoot[RO]{\footnotesize{\sffamily{1--\pageref{LastPage} ~\textbar  \hspace{2pt}\thepage}}}
\fancyfoot[LE]{\footnotesize{\sffamily{\thepage~\textbar\hspace{3.45cm} 1--\pageref{LastPage}}}}
\fancyhead{}
\renewcommand{\headrulewidth}{1pt} 
\renewcommand{\footrulewidth}{1pt}
\setlength{\arrayrulewidth}{1pt}
\setlength{\columnsep}{6.5mm}
\setlength\bibsep{1pt}

\twocolumn[
  \begin{@twocolumnfalse}
\noindent\LARGE{\textbf{Schematic Models for Active Nonlinear Microrheology}}%{{$^\dag$}}
\vspace{0.6cm}

\noindent\large{\textbf{M V Gnann,\textit{$^{a}$} I Gazuz,\textit{$^{a\ddag}$} A M Puertas,$^{\ast}$\textit{$^{b}$}, M Fuchs,\textit{$^{a}$} and
Th Voigtmann\textit{$^{a,c,d}$}}}\vspace{0.5cm}
%Please note that \ast indicates the corresponding author(s) but no footnote text is required. 

\noindent\textit{\small{\textbf{Received Xth XXXXXXXXXX 20XX, Accepted Xth XXXXXXXXX 20XX\newline
First published on the web Xth XXXXXXXXXX 200X}}}

\noindent \textbf{\small{DOI: 10.1039/b000000x}}
\vspace{0.6cm}
%Please do not change this text.

\noindent \normalsize{We analyze the nonlinear active microrheology of dense colloidal suspensions using a schematic model of mode-coupling theory. The model describes the strongly nonlinear behavior of the microscopic friction coefficient as a function of applied external force in terms of a delocalization transition. To probe this regime, we have performed Brownian dynamics simulations of a system of quasi-hard spheres. We also analyze experimental data on hard-sphere-like colloidal suspensions [Habdas \textit{et~al., Europhys.~Lett.}, 2004, \textbf{67}, 477]. The behavior at very large forces is addressed specifically.}
%{The abstract should be a single paragraph which summarises the content of the article. Any references in the abstract should be written out in full \textit{e.g.} [Surname \textit{et al., Journal Title}, 2000, \textbf{35}, 3523].}%
\vspace{0.5cm}
 \end{@twocolumnfalse}
  ]

\section{Introduction}
%Footnotes
%\footnotetext{\dag~Electronic Supplementary Information (ESI) available: [details of any supplementary information available should be included here]. See DOI: 10.1039/b000000x/}

%Please use \dag to cite the ESI in the main text of the article.
%If you article does not have ESI please remove the the \dag symbol from the title and the above footnotetext.

\footnotetext{\textit{$^{a}$~Fachbereich Physik, Universit\"at Konstanz, 78457 Konstanz, Germany. }}
\footnotetext{\textit{$^{b}$~Departamento de F\'\i{}sica Aplicada, Universidad de Almer\'\i{}a, 04.120 Almer\'\i{}a, Spain. E-mail: apuertas@ual.es}}
\footnotetext{\textit{$^{c}$~Institut f\"ur Materialphysik im Weltraum, Deutsches Zentrum f\"ur Luft- und Raumfahrt (DLR), 51170 K\"oln, Germany. }}
\footnotetext{\textit{$^{d}$~Zukunftskolleg, Universit\"at Konstanz, 78457 Konstanz, Germany. }}
\footnotetext{\ddag~\textit{Present address: Leibnitz Institute of Polymer Research Dresden, Hohe Stra\ss{}e 6, 01069 Dresden, Germany}}

%additional addresses can be cited as above using the lower-case letters, c, d, e... If all authors are from the same address, no letter is required

%\footnotetext{\ddag~Additional footnotes to the title and authors can be included \emph{e.g.}\ `Present address:' or `These authors contributed equally to this work' as above using the symbols: \ddag, \textsection, and \P. Please place the appropriate symbol next to the author's name and include a \texttt{\textbackslash footnotetext} entry in the the correct place in the list.}

%% MAIN TEXT BEGINS %%

Microrheology is a developing technique to acquire local information
on the viscous and elastic properties of complex fluids and soft matter.
\cite{Cicuta2007,Squires.2008,Kimura.2009,Squires.2010}
In active microrheology one uses a mesoscopic colloidal probe particle manipulated
by an external driving mechanism such as laser tweezers or magnetic forces,
and measures the response of that particle to the driving. The two major
complications in making sense of the obtained data are, firstly, that one
usually deals with suspending host liquids that have mesoscopic structure
on length scales comparable to the size of the probe, calling into question
simplifying assumptions treating the host liquid as a continuum.
\cite{Levine.2000}
Second, even
moderate external driving is sufficient to enter the nonlinear-response
regime of soft host liquids. On the other hand, if one is able to understand
the implications of this situation, the technique in principle gives access
to detailed information about structure-dynamics relationships of complex
soft matter. This makes active microrheology an ideal tool to probe, e.g.,
cellular environments. \cite{Cicuta2007,Lau.2003,Wilhelm.2008}

Here, we focus on the application of constant-forcing active microrheology to
colloidal glass formers, different from variable forcings such as
parabolic traps moving at constant velocity, \cite{Wilson2009} or where the
tracer moves at a constant speed. \cite{Sriram.2009,Sriram.2010}
Approaching the glass transition, the dynamics of
the host liquid becomes increasingly slow, heterogeneous, and stretched
over many orders of magnitude in time. Microrheology might, if properly
understood, provide valuable insights into the microscopic origins
of this slow structural relaxation dynamics.
\cite{Hastings.2003,Habdas.2004,Williams.2006}
The slow relaxation processes render highly nonlinear the
relation between the measured steady-state velocity of the probe and the
externally applied force. A pronounced superlinear rise in the velocity
is observed for forces far exceeding the scale set by thermal energy and
the particle size.
\cite{Habdas.2004} This force threshold can be interpreted as the strength
of nearest-neighbor cages that are broken by strong forcing before
they can relax due to structural relaxation.

We have recently presented a microscopic theory of nonlinear force-driven
active microrheology for dense colloidal suspensions: \cite{Gazuz.2009} based
on an integration-through transients (ITT) framework and mode-coupling
approximations, equations for the microscopic friction
coefficient $\zeta(F^\text{ex})$ were obtained, defined by the steady-state relationship
\begin{equation}\label{zetadef}
 \zeta\,\langle\vec v\rangle_{t\to\infty}=\vec F^\text{ex}\,,
\end{equation}
where angular brackets denote
the steady-state ensemble average reached at long times. These equations have a fully microscopic
foundation as they are based on the Smoluchowski equation for the colloidal
system (neglecting hydrodynamic interactions). They only require the equilibrium
static structure factor as input. This, however, makes them also rather
complicated to solve. In order to understand generic features of the
equations, we have devised schematic models, i.e., ad-hoc simplifications
of the original equations that are much easier to solve. The simplification
essentially amounts to dropping spatial information (wave-vector
dependences), at the cost of introducing a limited number of fitting
parameters that replace the static-structure-factor input.
One such schematic model was already presented earlier, \cite{Gazuz.2009}
but this model did not capture some features of the high-force limit
of the dynamics observed in experiment and simulation. In the following,
we present an extended model that levies this limitation by taking into
account separately the symmetry-breaking direction along the external force,
and directions perpendicular to this.
We are thus able to discuss the qualitatively different behavior of probe
fluctuations along those directions, and their effects on the friction in
the steady-state probe velocity.
The model is checked with computer simulations of a Brownian system,
which we perform, and experiments taken from Habdas et~al.\
\cite{Habdas.2004} The model correctly describes the qualitative features
of the friction coefficient as a function of density and external forcing,
and reproduces them quantitatively with a sensible choice of the parameters
for both the simulations and experiments.
The tracer position correlation function, a key quantity in
the theory, behaves similarly in simulation and theory.

\section{Computer Simulation}

We performed molecular-dynamics simulations for quasi-hard-sphere particles
governed by a Langevin equation, in the
following referred to as Brownian dynamics (BD) simulations as hydrodynamic
interactions are neglected. The system is
polydisperse to avoid crystallization (flat distribution of radii, with
half-width 10\% of the average radius), and the interaction potential is
given by $V(r) = k_BT (r/\sigma)^{-36}$, where $\sigma$ is the particles'
center-to-center distance. $k_BT=1$ and the average radius of a
host-liquid particle, $a=1$, set the units of energy and length.
All particles have the same mass, $m=1$.
The solvent friction coefficient, which also
sets the Langevin forces, is fixed to $\zeta_0=50$ in these units.
The near-equilibrium dynamics of
this system has been analyzed in detail before,
\cite{Voigtmann.2004b,Weysser.2010}
establishing a mode-coupling glass transition at an overall packing fraction
$\varphi\approx0.595$. At the highest packing fraction investigated here,
$\varphi=0.62$ the system cannot be equilibrated in the available computer
time; here we report measurements obtained after a waiting time
$t_w=2.5\times10^4$ after setting up the simulation runs. For forces exceeding
$F\approx35\,k_BT/a$, this is long enough to obtain results that do not depend
on $t_w$; for smaller forces, aging effects may still affect the data
at this highest packing fraction.

\begin{figure}
\centering
  \includegraphics[width=.95\linewidth]{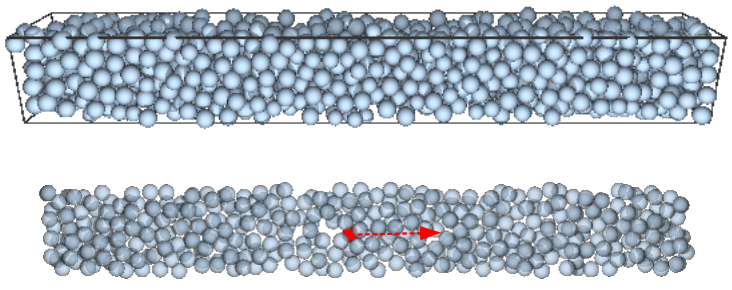}
  \caption{Snapshot of the simulated system at packing fraction
  $\varphi=0.55$, showing the box geometry and
  an examplary run with a randomly selected probe. An arrow indicates the
  direction of the applied force.
%\TODO{mf}{can/shall we submit videos?}
   }
  \label{snapshot}
\end{figure}

The simulation box is $8$ times longer in
the direction of the applied force than in the transversal directions, to allow
the study of longer trajectories of the tracer, and contains $1000$
particles, as shown in the snapshot in Fig. \ref{snapshot}. At 
the highest density studied here, $\varphi=0.62$, the width of the box is 
$\approx9.5a$, enough to relax the structural effects caused by the moving probe.
At time $t=0$, one particle is selected at random to be the probe particle
(hence, the probe radius $a_s=a=1$), and a constant external
force $\vec F^{\text{ex}}=\left( F^{\text{ex}},0,0 \right)$ is exerted over
it (in addition to the interparticle, friction and Brownian forces), and its
trajectory is monitored. When the
tracer has travelled a distance equal to half the box length, a new probe is
selected from scratch, and the average velocity 
is obtained from the average trajectory.
%\TODO{mf}{specify the number of samples}
The simulation scheme is identical to the one used recently. \cite{Gazuz.2009}
For small forces, trajectories were
recorded up to $t=10000$, before selecting a new probe. In these cases, sampling
was performed over $300$ trajectories obtained from different initial conditions
and tracer particles.

From the simulation runs we thus obtain the probe response in terms of the
friction coefficient $\zeta(F^\text{ex})$ at various packing fractions,
using the steady-state relationship, Eq.~\eqref{zetadef}.
We also analyze the dynamics of density fluctuations, 
characterized by the tagged-particle density correlation functions
$\phi^s_{\vec q}(t)=\langle\varrho^{s*}(\vec q,t)\varrho^s(\vec q)\rangle$.
Here, $\varrho^s(\vec q)=\exp[i\vec q\vec r_s]$ is the tagged-particle
density fluctuation to wave vector $\vec q$. In an isotropic homogeneous
liquid, the correlation function depends on the wave vector only through
$q=|\vec q|$; however, in the present context, application of the external
force degrades this spherical symmetry to a merely rotational one.
As above, angular brackets denote ensemble averaging; here, however
we distinguish equilibrium from steady-state averages, leading to,
respectively, transient and stationary correlation functions.

\section{Schematic Models}

In this section, we summarize the main equations defining the schematic
models.
Mode-coupling theory (MCT) in the integration-through transients framework
uses temporally and spatially resolved transient density correlation
functions to describe the nonlinear response of the system to the strong
external perturbation. In the case of constant-forcing active microrheology,
the transient tagged-particle density correlation
function $\phi^s_{\vec q}(t)$ is the basic dynamical quantity.
Acknowledging the symmetry-breaking direction of the external force,
it will be useful to distinguish for the probe particle the directions
$\vec q\parallel\vec F^\text{ex}$ and $\vec q\perp\vec F^\text{ex}$.
If $\vec q$ has a component parallel to the force, the displacement of
the particle due to the external force causes the correlator to become
complex valued, as it is the Fourier transform of a tagged-particle
density profile that is not centered on the origin. Formally, this is
a consequence of the non-Hermitian time-evolution operator in the
Smoluchowski equation without detailed balance. If $\vec q$ is perpendicular
to $\vec F^\text{ex}$, the correlator stays real. The internal forces
acting on the tracer are determined by tracer- and host-liquid density
fluctuations. Whitin MCT, the latter are captured by the intermediate
scattering function, $\phi_q(t)$, which, in an appropriate thermodynamic
limit can be taken as the equilibrium one
(being real and exhibiting full rotational symmetry).

The schematic model describes the dynamics of the system through
wave-number independent functions $\phi(t)$ and $\phi^s_\alpha(t)$ that
serve as proxies for the full correlation functions;
here, $\alpha\in\{\parallel,\perp\}$
tells apart a complex- and a real-valued tracer correlator.
The neglect of spatial resolution is motivated by equilibrium MCT,
where a factorization theorem states that all length scales are strongly
coupled and arrest simultaneously at the MCT transition. \cite{Goetze.2009}
The two kinds of tracer
correlators capture the dominant effects of the external force: some
correlators turn complex, while others remain (almost) real.
The non-Markovian equations of motion analogous to the microscopic
Mori-Zwanzig equations read
\begin{gather}\label{mz}
  \partial_t\phi(t)+\Gamma\left\{
  \phi(t)+\int_0^t m(t-t')\partial_{t'}\phi(t')\,dt'\right\}=0\,,\\
  \partial_t\phi^s_\alpha(t)+\omega^s_\alpha\left\{
  \phi^s_\alpha(t)
    +\int_0^t m^s_\alpha(t-t')\partial_{t'}\phi^s_\alpha(t')\,dt'\right\}=0\,.
\end{gather}
Here, $\Gamma$ models the short-time dynamics of the host-liquid particles;
we choose time units such that $\Gamma=1$. The short-time dynamics of
$\phi^s_\alpha(t)$ is affected by the non-Hermitian nature of the
underlying time-evolution operator and is modeled by
$\omega_\parallel^s=\Gamma_s(1-i\kappa_0F^\text{ex})$ and
$\omega_\perp^s=\Gamma_s$. The coefficient $\Gamma_s$ could be used to
describe a short-time diffusion coefficient of the probe that differs from
that of the host-liquid particles; we set $\Gamma_s=1$ for simplicity.
The strength of the external force relative to internal ones at short times is measured
by $\kappa_0$.

The memory kernels $m(t)$ and $m^s_\alpha(t)$ are approximated in MCT
as nonlinear functionals of the density correlators themselves. We take
for the schematic model
\begin{gather}
  m(t)=v_1\phi(t)+v_2\phi(t)^2\,,\label{mf12}\\
  m^s_\parallel(t)=\left[v_1^s\phi^{s*}_\parallel(t)\phi(t)
                  +v_2^s\phi^s_\perp(t)\phi(t)\right]
              /(1-i\kappa_\parallel F^\text{ex})\,,\label{mpar}\\
  m^s_\perp(t)=\left[v_1^s\phi^s_\perp(t)\phi(t)
              +v_2^s\Re\phi^s_\parallel(t)\phi(t)\right]
              /\left(1+\left(\kappa_\perp F^\text{ex}\right)^2\right)\,.\label{mperp}
\end{gather}
All coupling parameters $v_i$ and $v_i^s$ are taken to be real and
positive; they model force-free equilibrium structural correlations.
Equation \eqref{mf12} just specifies the well-known $\text{F}_\text{12}$ model
that is regularly used to analyze linear-response dynamics of glass forming
liquids within mode-coupling theory. \cite{Goetze.2009}
The quadratic polynomial mimics the feedback mechanism termed `cage effect'
that causes the slowing down in the structural relaxation of the host
liquid.
The bilinear coupling to $\phi^s_\alpha(t)$ and $\phi(t)$
in Eqs.~\eqref{mpar} and \eqref{mperp}
mimics that internal forces on the probe relax via its motion and via
rearrangements
of the surrounding particles. Both tracer motion
parallel and perpendicular to the external force can relax the local
friction.
Spatial inversion symmetry demands that $\phi^s_\parallel(t)$ and
$\phi^{s*}_\parallel(t)$ are coupled, which is assured by the complex
conjugate appearing in $m^s_\parallel(t)$.
The ensuing force driven delocalization
transition can be mapped out using the external force as single relevant control parameter. \cite{Gazuz.2009}
 Since $m^s_\perp(t)$ is
real-valued, only the real part $\Re\phi^s_\parallel(t)$ can enter
in Eq.~\eqref{mperp}.
The external force enters differently in
$m^s_\parallel(t)$ and $m^s_\perp(t)$: the kernel $m^s_\parallel(t)$ is
complex-valued, and $F^\text{ex}$ suppresses both parts, while $m^s_\perp(t)$
needs to stay real-valued.
Including $F^\text{ex}$ in the denominator ensures the correct physical
behavior that large forces decrease the coupling strength.
Equations \eqref{mpar} and \eqref{mperp} explicitly express the symmetry
under inversion of the force: $\phi^s_\parallel(-F)=\phi^{s*}_\parallel(F)$
but $\phi^s_\perp(-F)=\phi^s_\perp(F)$.
The positive parameters
$\kappa_\parallel$ and $\kappa_\perp$ measure the effective force in
the directions dominantly parallel and dominantly perpendicular to the
external force. They should, by analogy to the microscopic model, be
smooth functions of the thermodynamic control variables, \cite{Gnann.2009}
and we take for simplicity $\kappa_\parallel=\kappa_0$ which was the
choice in the original schematic model. \cite{Gazuz.2009}

Finally, the friction coefficient is expressed through dynamical correlation
functions via a nonequilibrium generalization of the Green-Kubo relation
central to the ITT framework.
\cite{Gazuz.2009} Following a mode-coupling approximation, we arrive at,
schematically, $\zeta=\zeta_0+\Delta\zeta$,
\begin{equation}\label{zetaschem}
  \Delta\zeta/\zeta_0 =
   \mu_\parallel\, \Gamma_s\int_0^\infty\!\!\! dt\,\Re\phi^s_\parallel(t)\phi(t)
  +\mu_\perp\,\Gamma_s\int_0^\infty\!\!\! dt\,\phi^s_\perp(t)\phi(t)\,,
\end{equation}
where the parameters $\mu_\alpha\ge0$ replace angular-dependent coupling
coefficients that
are given by the equilibrium structure functions in the microscopic theory.
Again this MCT approximation expresses that the friction on the
probe arises from the cumulated transient fluctuations of probe and
host-fluid densities. From the correlation function $\alpha=\parallel$,
only the real part enters due to spatial inversion symmetry.
Equations \eqref{mz} to \eqref{zetaschem} specify our schematic model
completely. They can further be motivated by an ad-hoc simplification
of the full microscopic MCT, restricting wave-vector integrals to
four wave vectors, $\vec q\parallel\vec F^\text{ex}$ and
$\vec q\perp\vec F^\text{ex}$. \cite{Gnann.2009}
For each $\vec q$ also $-\vec q$
has to be taken to obey the required symmetry
$\phi^s_{-\vec q}(t)=\phi^{s*}_{\vec q}(t)$; this justifies the
appearance of the complex conjugate in \eqref{mpar}.
%the memory kernel $m^s_\parallel(t)$.

The earlier schematic model for active microrheology \cite{Gazuz.2009}
only included a single complex correlator, concentrating on the novel
transition that a particle is pulled mobile by a finite external force.
The present three-correlator model
reduces to this two-correlator model upon setting $v_2^s=\mu_\perp=0$.
Also, for $F^\text{ex}=0$, we obtain $\phi^s_\parallel(t)=\phi^s_\perp(t)$
and the model reduces to the Sj\"ogren model
of tagged-particle dynamics in the near-equilibrium glass forming liquid,
with a single probe-coupling parameter $v^s=v_1^s+v_2^s$; the latter is
a standard model for analysis of linear-response dynamics of glass forming
liquids, e.g., in terms of scattering spectra. \cite{Goetze.2009}

The present model additionally captures a nontrivial limit
of the friction at high force, $\Delta\zeta_\infty\equiv
\Delta\zeta(F^\text{ex}\to\infty)$. The model with $\mu_\perp=0$
predicts $\Delta\zeta_\infty=0$ as
the presence of the term $\propto iF^\text{ex}$
in $\omega^s_\parallel$ leads to arbitrarily fast oscillations in
$\phi^s_\parallel(t)$; in the integral determining $\Delta\zeta$ these
lead to cancellations such that the overall integral vanishes.
Only when $\mu_\perp\neq0$ does one get an additional contribution to
$\Delta\zeta$ in the large force limit.
If we furthermore neglect all memory kernels, we obtain as the
low-density (weak-coupling) limit of the schematic model the ratio
$\Delta\zeta_\infty/\Delta\zeta(F^\text{ex}\to0)=
\mu_\perp/(\mu_\parallel+\mu_\perp)$.
This ratio is known to equal $1/2$ exactly in the low-density limit
\cite{Squires.2005}, checked also in our simulations. We thus, for simplicity,
set $\mu_\parallel=\mu_\perp=\mu$ throughout in the following.

\section{Data Analysis}

The schematic model presented above has several fit parameters. Two of them,
$v_1$ and $v_2$, do not depend on the probe at all; they specify the
dynamics of the host liquid and its vicinity to the glass transition
which occurs for some critical coupling $(v_1^c,v_2^c)$. Thus, these
two parameters can in principle be determined independently. In the cases
we analyze, the only physical control parameter is the packing fraction
of the suspension, $\varphi$.
One demands that the fit parameters follow a linear relationship
$(v_1,v_2)=(v_1^c,v_2^c)(1+\epsilon)$ with $\epsilon\propto(\varphi-\varphi^c)/
\varphi^c$, where $\varphi^c$ is the glass-transition packing fraction of
the suspension. The critical point is then fixed by demanding the
MCT exponents of the schematic model to match those found in experiment
or simulation. In practice, this fixes the pair $(v_1^c,v_2^c)$, and
one fits $\epsilon$ for each packing fraction, biasing it to follow
the stated linear relationship.

Two parameters $v_1^s$ and $v_2^s$ describe the coupling strength of the
probe to the host suspension; their sum $v^s$ could be determined from
analyzing linear-response measurements independently. In principle, they depend
on the density of the host; we will in the following thus allow $v^s$ to
increase with increasing packing fraction. Inspection of the microscopic
model and its symmetries \cite{Gnann.2009} suggests the fixed ratio
$v^s_1/v^s_2=2$. The remaining parameters $\kappa_\perp$ and $\kappa_\parallel$
are global parameters (fixed differently for the simulation and for the
experiment to reflect the difference in systems) specifying the
relative influence of the external force on the memory kernel.

In our fits of both BD simulation and experimental data, we aimed at keeping
as many parameters fixed as is physically plausible. First, we set
$v_2^c=2$, resulting in $v_1^c=2(\sqrt2-1)$, which renders the near-equilibrium
asymptotic behavior of the schematic model close to that observed for
the hard-sphere glass transition. Furthermore, $\epsilon(\varphi)$ is
estimated from the known linear-response regime for the simulation data;
for the fits to the experimental data we require that the obtained
values of $\epsilon$ are of the same magnitude for $\varphi$ not too close
to the critical point.
To capture the density dependence of the high-force plateau, we also increase
$\mu$ with increasing density.

\begin{figure}[t]
\centering
  \includegraphics[width=.92\linewidth]{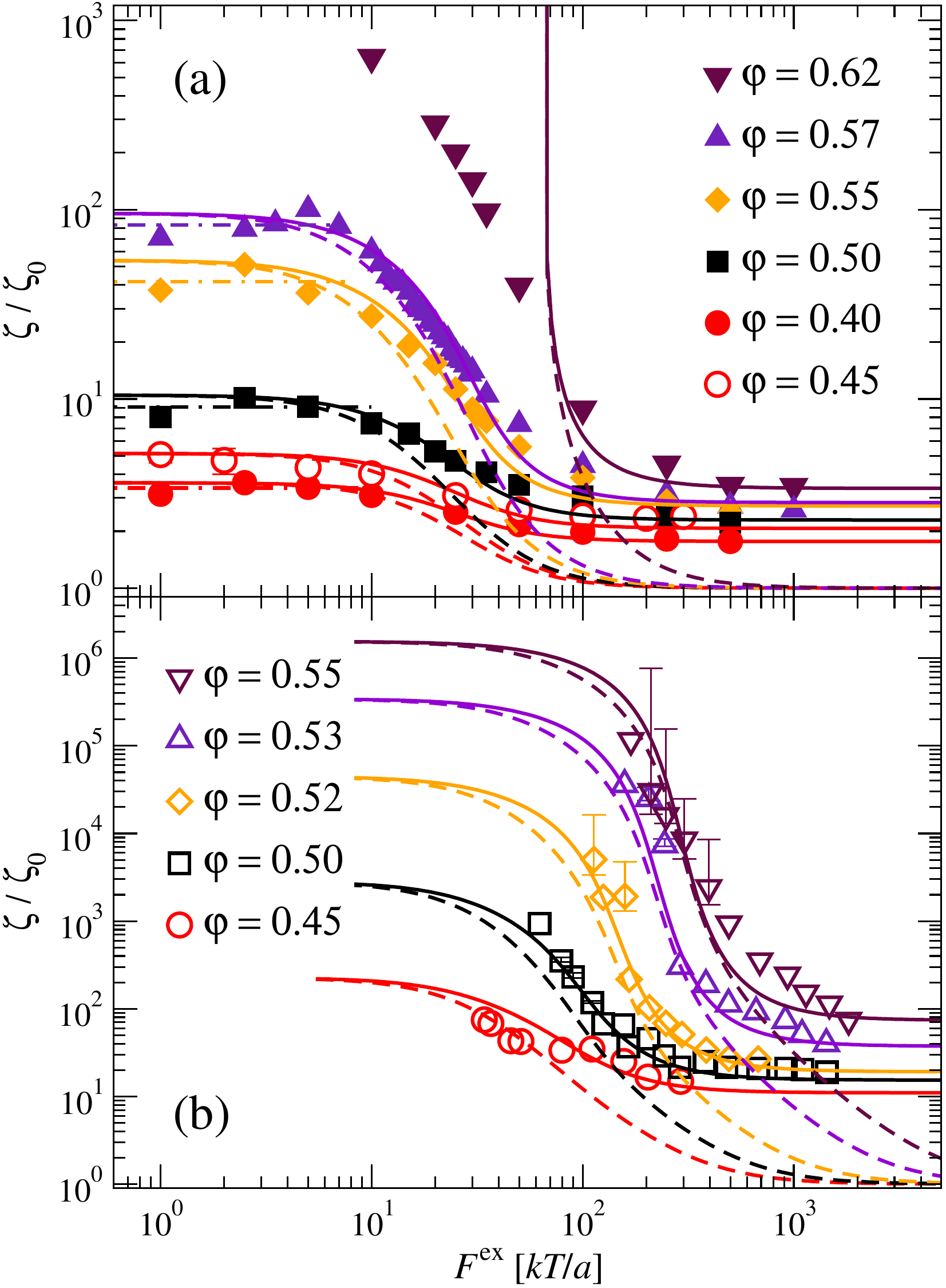}
  \caption{Microscopic friction coefficients $\zeta$ as a function of
    external force $F^\text{ex}$: symbols are (a) Brownian dynamics
    (filled) and hard-sphere simulations of Ref.~\citenum{Carpen.2005}
    (open); (b) experiment of Ref.~\citenum{Habdas.2004}, evaluted from
    the original velocity data (Fig.~\protect\ref{fgr:fits-vel}) and
    translated to dimensionless quantities by setting
    $\zeta_0=0.095\,\text{mg/s}$ and $1\,\text{pN}=271.9\,k_BT/a$.
    Lines are fits with schematic models of
    mode-coupling theory, taking only fluctuations in direction of the force
    into account (dashed), or also in the perpendicular direction (solid
    lines). Horizontal dot-dashed bars in (a) give the quiescent self-diffusion coefficients.
   }
  \label{fgr:fits-zeta}
\end{figure}

We first turn to a discussion of the friction coefficient $\zeta$, shown
in Fig.~\ref{fgr:fits-zeta} for our Brownian dynamics simulations, and
for the experiment by Habdas et~al.\cite{Habdas.2004}. We have
translated the latter from the reported velocity data and into dimensionless
units by using the known solvent properties, leading to
$\zeta_0=0.095\,\text{pN}\text{s}/\text{$\mu$m}$. In
Fig.~\ref{fgr:fits-zeta}, also a result from simulations of monodisperse
hard spheres by Carpen and Brady \cite{Carpen.2005}
is shown, highlighting that our choice of soft-sphere
potential and polydispersity are not crucial for the present discussion.
While in both simulations, probe and host-liquid particles were of the
same size, $\delta=a_s/a=1$, the quoted values for the experiment in
Ref.~\citenum{Habdas.2004} are $a_s=2.25\,\text{$\mu$m}$ and
$a=1.1\,\text{$\mu$m}$, leading to a size ratio of $\delta\approx2.05$.
This difference explains a shift in force scales when comparing the
two sets of data.

Solid lines in Fig.~\ref{fgr:fits-zeta} are fits using the
schematic model.
%; for the experimental data, the fitted quantity is in fact
%the steady-state velocity shown in Fig.~\ref{fgr:fits-vel}.
Fit parameters are given in Table~\ref{tbl:fitparameters}; additionally,
under the constraint that the fits describe reasonably the
known (simulation) or anticipated (experiment) linear response,
we found $(\kappa_\parallel,\kappa_\perp)=(0.25,0.0625)a/k_BT$ for the
simulation, and $(\kappa_\parallel,\kappa_\perp)=(0.046,0.0115)a/k_BT$
for the experimental data to give satisfying results.
These are not the only parameter sets that give reasonable fits; on the
schematic level, no more physical significance can be attached to the choice of
parameters.

As is evident, the model reproduces the qualitative features of the
available data, in particular the steep change of $\zeta$ already in the
liquid around a
threshold of ${\mathcal O}(20\,k_BT/a)$ for the simulation
and ${\mathcal O}(200\,k_BT/a)$ for the experiment (larger
due to the larger force required to pull free a larger probe).
As pointed out earlier, \cite{Gazuz.2009}
this threshold force is in the schematic model precisely defined as the
force that is needed to pull the probe particle free even if the suspending
host is glassy. Reassuringly, microscopic calculations for the hard-sphere
model yielded values comparable to the ones found in our simulation.
From the schematic model, we get for the critical force just at the
glass transition $F^\text{ex}_c\approx63\,k_BT/a$ for the
simulation, and $F^\text{ex}_c\approx800\,k_BT/a\approx3\,\text{pN}$ for the experiment. At higher densities, even larger forces are needed to break
the cages.
According to MCT, the host system is in an ideal-glass state for
densities $\varphi>\varphi^c\approx0.595$ in the simulation; hence, at
vanishing external force, the single-particle friction coefficient is
infinite since the tracer remains localized in the glass. The simulation
data for $\varphi=0.62$ still show finite values of $\zeta$ at all forces, which may partly
be due to aging effects;
it may also reflect deviations of the simulated glass from the idealized
MCT description.

\begin{figure}[t]
\centering
  \includegraphics[width=.92\linewidth]{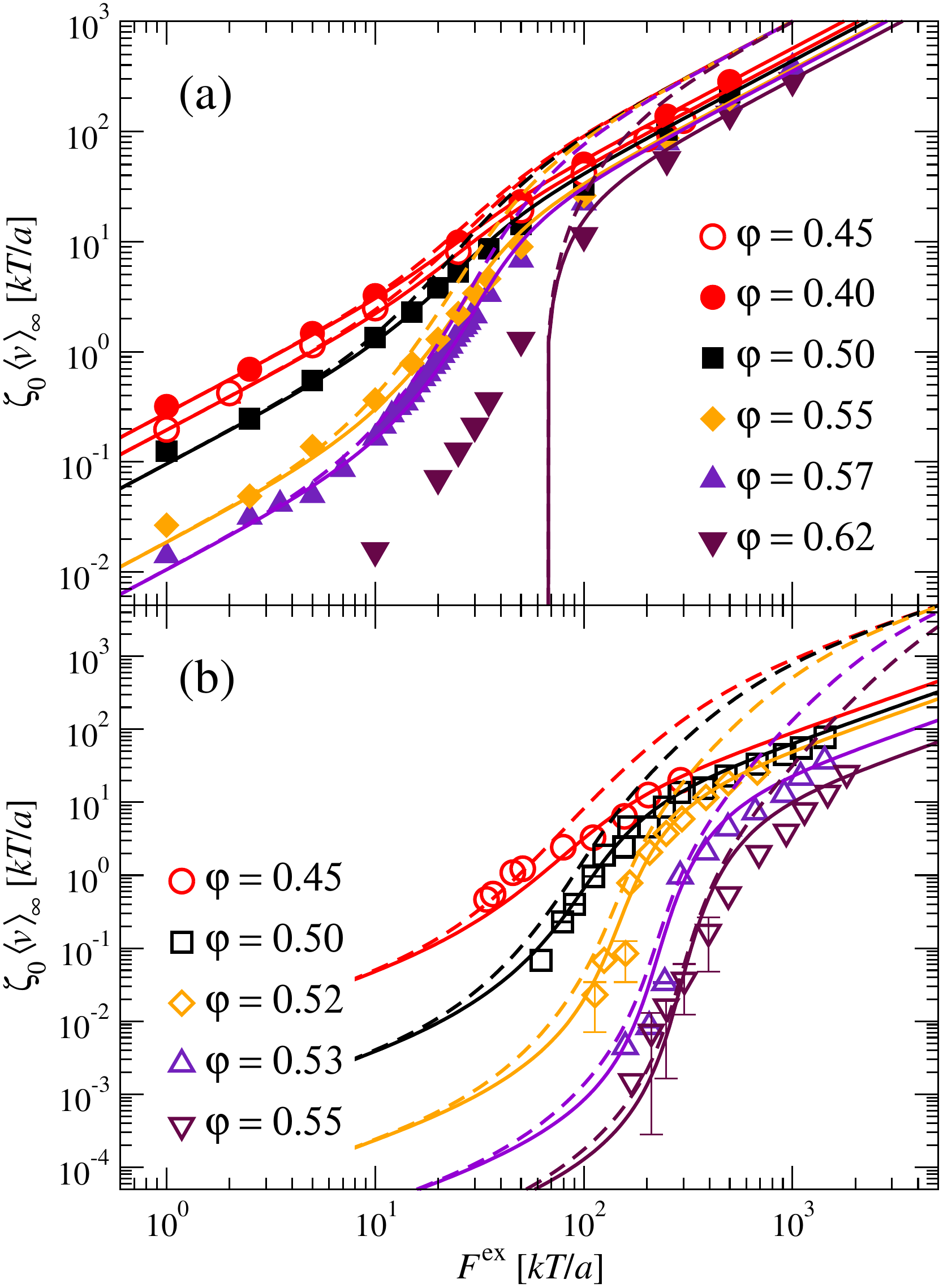}
  \caption{Steady-state probe velocities $\langle\vec v\rangle_\infty$
    corresponding to Fig.~\protect\ref{fgr:fits-zeta},
    as a function of
    external force: computer-simulation (a) and
    experiment (b).
    Lines are the same schematic-model fits
    as in Fig.~\protect\ref{fgr:fits-zeta}.}
  \label{fgr:fits-vel}
\end{figure}

Before turning to a discussion of the large-force behavior, let us
briefly discuss the steady-state probe velocities corresponding to
Fig.~\ref{fgr:fits-zeta}. This is the quantity originally obtained in
the experiment, and is shown in Fig.~\ref{fgr:fits-vel}.
The plot highlights the difficulty to stay within the linear-response
regime in dense colloids. Velocities smaller than $10^{-3}\,k_BT/a\zeta_0
\sim10^{-3}\,D_0/a$ have to be measured, where $D_0\sim\text{$\mu$m}^2/\text{s}$
is a typical short-time diffusion coefficient for a colloidal particle
of radius $a\sim1\,\text{$\mu$m}$. The resulting velocities are
on the order of a few particle diameters per week,
which clearly is a challenge. Such small velocities pose the additional
problem of ensuring a typical enough sample of the configuration space
required in the ensemble average, a problem also for the simulations.

In both Figs.~\ref{fgr:fits-zeta} and \ref{fgr:fits-vel}, dashed lines
represent the predictions of the schematic model when setting $\mu_\perp=0$,
i.e., ignoring the contribution of fluctuations perpendicular to the
external force to the friction coefficient. Fits with somewhat different
parameters also setting $v_2^s=0$ have been presented earlier;
\cite{Gazuz.2009} they are qualitatively the same.
The resulting curves still
show the generic features of the rapid decrease in friction in the
vicinity of the force threshold.
The most striking difference is that for $\mu_\perp=0$, one obtains
$\zeta(F^\text{ex}\to\infty)=\zeta_0$, the free-particle friction
coefficient, as explained above.
This behavior is not supported by the data, which clearly
show a larger, density-dependent high-force plateau in $\zeta(F^\text{ex})$.

At least within the schematic model, including fluctuations perpendicular to
the direction of the external force restores the nontrivial high-force
plateau. Here, $\phi^s_\perp(t)$ remains a real-valued positive function
whose decay is fixed by the equilibrium short-time relaxation of the
probe particle. Consequently, a finite contribution remains for
$\Delta\zeta$ even as $F^\text{ex}\to\infty$.

\begin{table}[b]
\small
  \caption{\ Fit parameters for the MCT schematic model}
  \label{tbl:fitparameters}
  \begin{tabular*}{.5\textwidth}{@{\extracolsep{\fill}}lllll}
    \hline
                      & $\epsilon$ & $v_s $ & $\mu$ \\
    \hline
    BD $\varphi=0.40$ & -0.98 & 9 & 0.45 \\
    HS $\varphi=0.45$ & -0.87 & 9 & 0.59 \\
    BD $\varphi=0.50$ & -0.54 & 9.75 & 0.62 \\
    BD $\varphi=0.55$ & -0.23 & 11.25 & 0.74 \\
    BD $\varphi=0.57$ & -0.185 & 14.25 & 0.78 \\
    BD $\varphi=0.62$ & 0.185 & 33 & 0.93\\
    \hline
    exp $\varphi=0.45$ & -0.17 & 4.5 & 14.3 \\
    exp $\varphi=0.50$ & -0.07 & 7.5 & 20 \\
    exp $\varphi=0.52$ & -0.03 & 15 & 25 \\
    exp $\varphi=0.53$ & -0.02 & 30 & 50 \\
    exp $\varphi=0.55$ & -0.015 & 45 & 100 \\
    \hline
  \end{tabular*}
\end{table}

\begin{figure}
\centering
  \includegraphics[width=.9\linewidth]{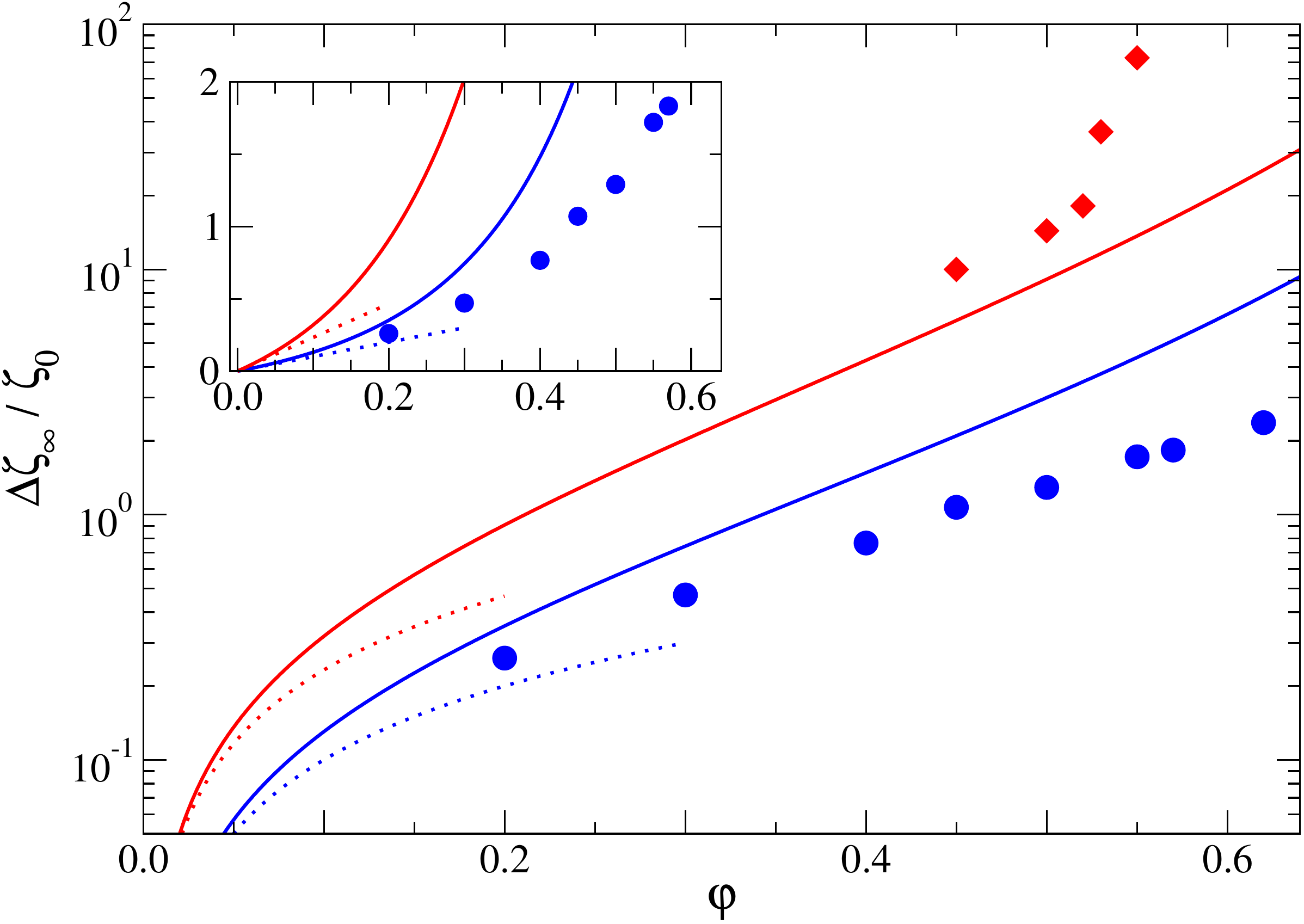}
  \caption{High-force limit of the microscopic friction increment,
  $\Delta\zeta_\infty/\zeta_0
  =\zeta(F^\text{ex}\to\infty)/\zeta_0-1$, obtained
  from simulation (circles; size ratio $\delta=a_s/a=1$) and experiment
  (squares; $\delta=2.05$), as a function of packing fraction
  $\varphi$. Lines are the low-density estimates (dotted) by
  Squires and Brady,\protect\cite{Squires.2005} corrected for packing
  effects by the equilibrium contact value from the BMCSL equation of
  state for these size ratios (solid lines).}
  \label{fgr:zetainf}
\end{figure}

A comparison between simulation and experiment regarding the high-force
plateau in $\zeta(F^\text{ex})$ is instructive. Replotting the experimental
data as in Fig.~\ref{fgr:fits-zeta} reveals that the data is not yet in
the limiting regime $F^\text{ex}\to\infty$ for all densities, but our schematic-model
fits suggest a confident extrapolation.
We obtain the values shown in Fig.~\ref{fgr:zetainf} as symbols. The
difference between the two data sets is the result of two physical differences
in the systems: the ratio $\delta=a_s/a$ and
the influence of hydrodynamic interactions. To demonstrate this, we have
included in Fig.~\ref{fgr:zetainf} as dotted lines the result obtained by
Squires and Brady. \cite{Squires.2005} Evaluating the hydrodynamic equations
for $\varphi\to0$, they obtained
$\Delta\zeta_\infty/\zeta_0=(1/4)\varphi(1+\delta)^2$.
%The simulation data indeed approaches this low-density asymptote (shown dotted)
%at small density, but deviates already around $\varphi=0.2$.
A first-order account for packing effects
in the host liquid is obtained by replacing the packing fraction $\varphi$
in this expression by $\varphi g_d(\varphi)$, where $g_d(\varphi)$ is the
probe--host-particle contact value. For the latter, we use the form
corresponding to the BMCSL equation of state for hard-sphere mixtures
due to Grundke and Henderson, \cite{Grundke.1972} in the limit
of vanishing probe-particle density,
$g_d(\varphi)=(1-\varphi+3\varphi\delta/(1+\delta))/(1-\varphi)^2
+2\varphi^2(\delta/(1+\delta))^2/(1-\varphi)^3$. Note that this expression
differs slightly from the more common Carnahan-Starling result for
$\delta=1$, used, e.g., by Carpen and Brady. \cite{Carpen.2005}
As shown in Fig.~\ref{fgr:zetainf}, this theory describes reasonably well the
experimental data when we set $\delta=2.05$. The increase
due to size-ratio effects alone is about a factor $4$ at $\varphi\approx0.5$.
The contact-value-corrected low-density expression for $\delta=1$ exceeds
the BD results; the difference might be due to the polydispersity
and the slightly soft potential used in the simulations, as both factors
reduce $g_d(\varphi)$. At $\varphi<0.2$, the simulation data
approaches the low-density asymptote.

It is worth pointing out that the friction coefficient at small forces,
$F^\text{ex}\to0$,
shows completely different scaling than the one for $F^\text{ex}\to\infty$:
while the latter is well described by hydrodynamics corrected for pair-density
effects, the former is governed by the approach of the equilibrium system
to the glass transition, where $\zeta(F^\text{ex}\to0)\to\infty$ as
$\varphi\to\varphi^c$ from the liquid side. This is clearly demonstrated
by the data shown in Fig.~\ref{fgr:fits-zeta}. We have checked that at
low densities, $\Delta\zeta(F^\text{ex}\to0)/\Delta\zeta(F^\text{ex}\to\infty)
=2$ as predicted \cite{Squires.2005,Carpen.2005} also holds for our
simulations (up to about $\varphi\approx0.2$).
Clearly, at higher density $\Delta\zeta(F^\text{ex}\to0)$
rises much more dramatically than the $F^\text{ex}\to\infty$
friction coefficient.

\begin{figure}
\centering
  \includegraphics[width=.95\linewidth]{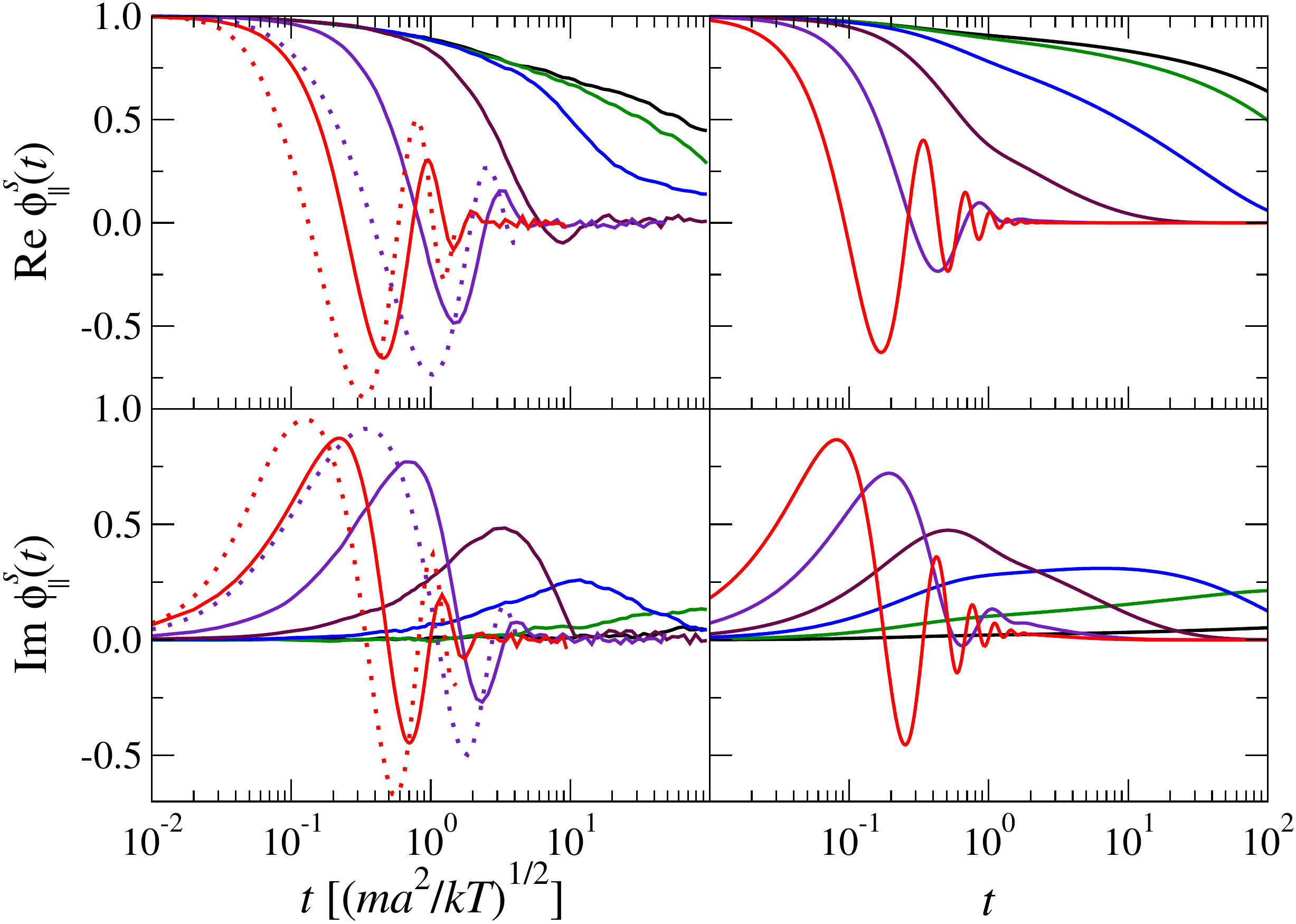}
  \caption{Probe-density correlation functions $\phi^s_\parallel(t)$ from
    BD simulations at $q=3.85 / a$ and $\varphi=0.55$ (left) and from the
    corresponding schematic model fit (right).
    Top (bottom) panels show the real (imaginary)
    part of the correlators. For the simulation,
    $a F^\text{ex}/(k_B T) = 1, 5, 15, 35, 100,$ and $250$ (right to
    left).}
  \label{fgr:phipar}
\end{figure}

\begin{figure}
\centering
  \includegraphics[width=.95\linewidth]{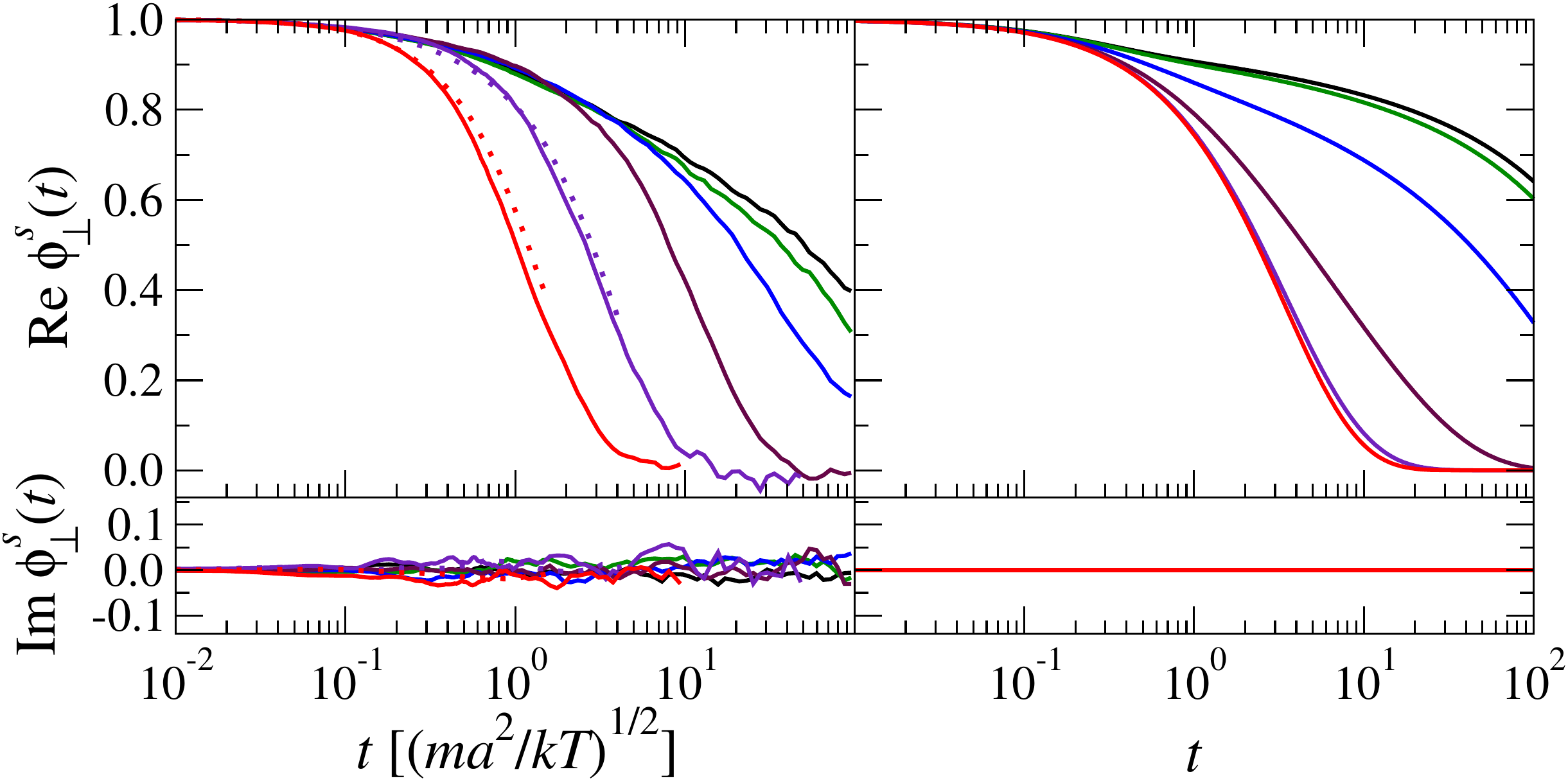}
  \caption{Probe-density correlation functions $\phi^s_\perp(t)$ from
    BD simulations at $q=3.85 / a$ and $\varphi=0.55$ (left) and from the
    corresponding schematic model fit (right).
    Top (bottom) panels show real (imaginary) parts
    of the correlators for forces as in Fig.~\protect\ref{fgr:phipar}.}
  \label{fgr:phiperp}
\end{figure}

The strength of the MCT model is that it can not only explain the available
data for the friction coefficient or steady-state velocity, but also the
underlying probe dynamics in terms of the tagged-particle density correlation
functions. We have obtained both the stationary and, for some values of
$F^\text{ex}$, also the transient correlation
functions from the BD simulation; both are shown in the left-hand panels
of Figs.~\ref{fgr:phipar} and \ref{fgr:phiperp} as continuous and
dotted lines, respectively, for a wave number $qa=3.85$,
packing fraction $\varphi=0.55$ and various external forces.
In Fig.~\ref{fgr:phipar}, the alignment
$\vec q\parallel\vec F^\text{ex}$ has been chosen,
while Fig.~\ref{fgr:phiperp} shows $\vec q\perp\vec F^\text{ex}$.
Even at
the largest forces we investigate, the difference between stationary and
transient correlation functions is not qualitative.
In the direction of the force, the transient correlation function
decays faster than the stationary one. We interpret this as being due to
a ``wave front'', i.e., a region of increased density building up in front
of the pulled particle, which slows down the structural rearrangement in
this direction once it is established in the steady state (see also
the discussion in Ref.~\citenum{Squires.2005}). Just after applying the
external force, this wave front has not been built up for short times,
and this is probed by the transient correlation function.

We now focus on the stationary correlation functions that are easier
to obtain from the simulation, since the difference to the transient
quantity is not important for our purposes.
In mode-coupling
theory based on integration through transients, the transient correlation
functions are the central quantities, and the evolution towards the
stationary functions needs to be understood separately. \cite{Krueger.2010}
Thus, the schematic-model correlators that are
shown in the right-hand panels should, strictly speaking, be interpreted
as transient correlation functions.

One recognizes a number of generic features in the correlation functions,
that are present in the simulation and correctly reproduced by our
schematic model. First, $\phi^s_\parallel(t)$ is a complex quantity, and it
shows pronounced oscillations at large $F^\text{ex}$ that are indicative
of a probe moving with a finite velocity that is faster than the
diffusive exploration of its configuration space. \cite{Gazuz.2009}
For small external forces, the correlation functions stay close to the
equilibrium tagged-particle correlator recovered for $F^\text{ex}\to0$,
indicating the linear-response regime. As $F^\text{ex}$ is increased,
the slow relaxation of the correlator is increasingly accelerated,
which is the microscopic origin for the pronounced force thinning observed
in Figs.~\ref{fgr:fits-zeta} and \ref{fgr:fits-vel}.

Taking $\vec q\perp\vec F^\text{ex}$, one instead recovers a real-valued
correlation function. Figure~\ref{fgr:phiperp} confirms this prediction
for the BD computer simulation: the calculated imaginary part is zero within
the noise level. Here, too, large forces lead to a shortening of the
equilibrium slow relaxation time, but no oscillations are seen, as the
probe particle does not perform net motion in the direction perpendicular
to the force. At very high $F^\text{ex}$, the schematic model predicts
a saturation, where $\phi^s_\perp(t)$ approaches an exponential decay on
the short-time diffusion scale. In the simulation, this is not (yet) clear.
The integral over this remaining force-independent decay is, in the
schematic model, the cause for the non-vanishing high-force friction
increment over the solvent friction. The model thus predicts the
large-force friction to be dominated by details of the short-time
motion, but not by the slow structural relaxation. The good quality
of the low-density predictions for $\Delta\zeta_\infty$ shown in
Fig.~\ref{fgr:zetainf} supports this.

\begin{figure}
\centering
  \includegraphics[width=.9\linewidth]{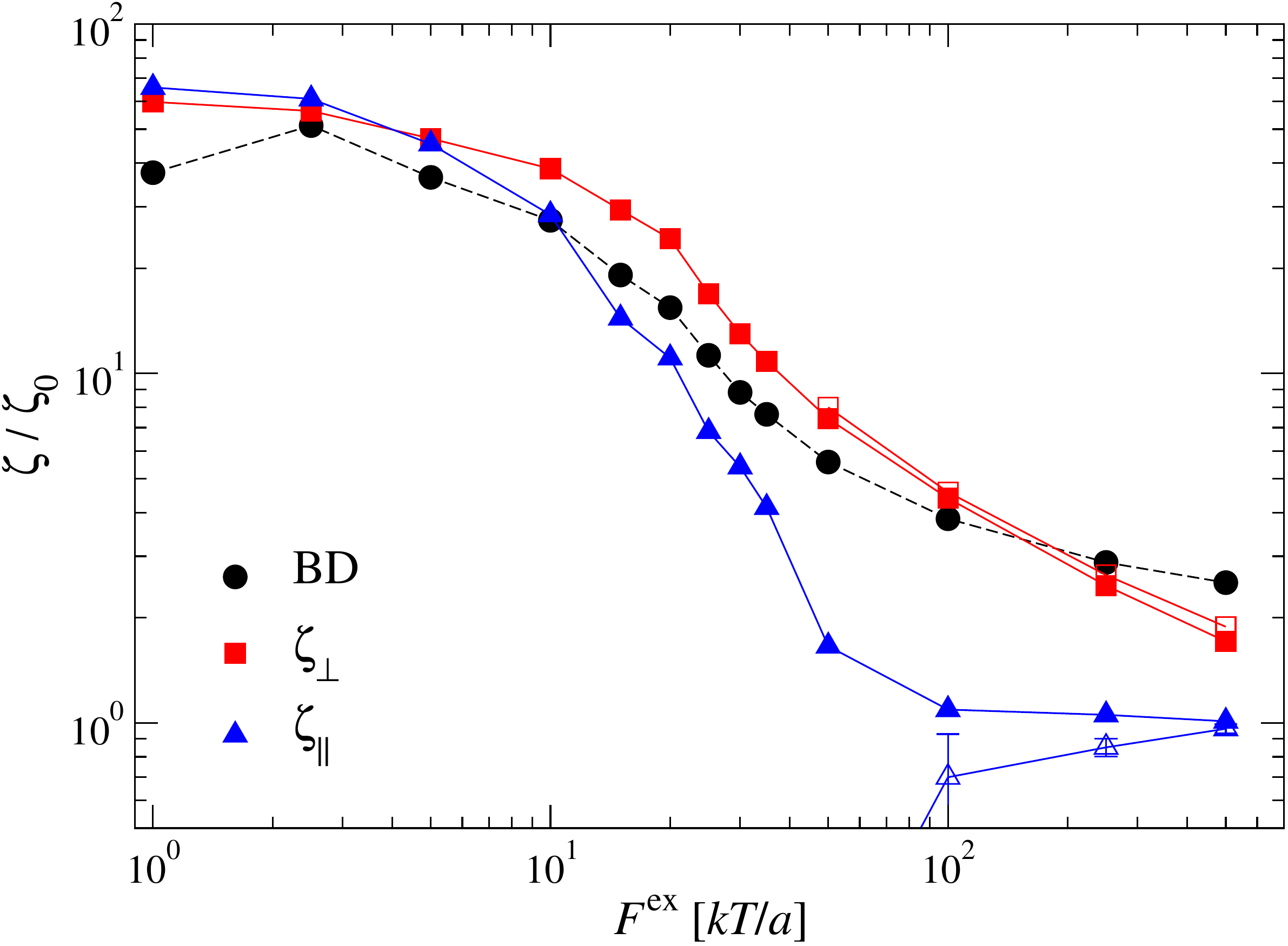}
  \caption{Microscopic friction $1+\Delta\zeta(F^\text{ex})/\zeta_0$
    for the BD simulation at $\varphi=0.55$. Circles: simulation data
    from Fig.~\protect\ref{fgr:fits-zeta}; squares (triangles):
    contributions in Eq.~\protect\eqref{zetadef}
    coming only from the simulated $\vec q\perp\vec F^\text{ex}$
    ($\vec q\parallel\vec F^\text{ex}$) correlation functions shown
    in Figs.~\protect\ref{fgr:phipar} and \protect\ref{fgr:phiperp}.
    Filled (open) symbols use the stationary (transient) correlation
    functions.
  }
  \label{fgr:numint}
\end{figure}

Further support for the schematic-model interpretation assigning
distinct roles to the principal directions $\vec q\perp\vec F^\text{ex}$
and $\vec q\parallel\vec F^\text{ex}$ at large external forces comes from
the simulation data in Figs.~\ref{fgr:phipar} and \ref{fgr:phiperp}.
In the microscopic theory, Eq.~\eqref{zetaschem} is replaced by a
wave-vector dependent integral over $\phi^{s*}_{\vec q}(t)\phi_q(t)$, taking
into account fluctuations in all directions relative to the force.
In principle, this integral could be evaluated with simulation data
for the (transient) correlation functions, without making use of the
MCT equations of motion. It is a formidable task to sample enough
correlation functions with sufficient statistics in the simulation.
To give a qualitative impression,
we only evaluate the integral over the simulated correlators for the two particular
wave vectors shown above separately; this corresponds to the two
contributions in the schematic Eq.~\eqref{zetaschem}. The host-liquid
correlation function needed in this equation has also been evaluated
at the same wave number, $q=3.85/a$, in the BD simulation.
The result is shown in Fig.~\ref{fgr:numint} for an exemplary packing fraction.
The integral using $\vec q\parallel\vec F^\text{ex}$ (triangle symbols)
gives $\zeta\approx\zeta_0$ at large forces, i.e.\
$\Delta\zeta(F^\text{ex})\to0$. Also in the BD simulation, the oscillations
in $\phi^s_{\vec q}(t)$ in the direction parallel to the external force
cancel each other, and the main contribution to the friction at high
external forces stems from density fluctuations to wave vectors perpendicular
to the force (squares in the figure). Our data for the transient correlation
functions is not of sufficient quality to be used in the integration,
although this should strictly speaking be done. For the highest forces
we simulated, we give a numerical estimate for the integral value
according to Eq.~\eqref{zetaschem} using the transient correlators
(open symbols in the figure).
The difference is not large (slightly negative values in some cases indicate
the magnitude of the error in the integration procedure, due to cutting
off at too early times). This further corroborates the point, that
wave-vector contributions perpendicular to the external force are
prominent in the large-force limit.

\section{Conclusions}

We have presented a schematic mode-coupling-theory analysis of
active nonlinear microrheology investigating the
dynamics of a probe particle pulled by a strong constant force
through a glass-forming dense colloidal suspension.
Extending previous analysis, \cite{Gazuz.2009} our schematic model
captures the qualitatively distinct behavior of probe-density correlation
functions for fluctuations in the direction of and perpendicular to the
external force. The distinct features -- complex-valued and at high force
oscillating correlation functions in the direction of the force, in contrast
to force-thinning real-valued functions perpendicular to $\vec F^\text{ex}$
-- reflect the property that the probe-particle density distribution is
biased in the direction of the force and attains a finite steady-state
velocity superseeding the quiescent motion at high enough forces, while
perpendicular to the force, no net motion occurs. These
features are confirmed in Brownian dynamics simulations.

Within the schematic model, contributions to the microscopic friction
coefficient $\zeta(F^\text{ex})$ coming from the different directions start
being distinguishable upon entering the nonlinear regime. Both contributions
drop sharply around a threshold that is large compared to forces induced
by thermal fluctuations alone. They lead to pronounced force thinning
and the eventual delocalization of the driven probe even from a glassy
surrounding matrix, at a critical force $F^\text{ex}_c$. The contribution
to the friction from fluctuations parallel to the external force vanishes
as $F^\text{ex}\to\infty$, but the contribution from the perpendicular
directions remains finite, giving rise to a non-trivial high-force plateau
above the Stokesian pure-solvent value.
The existence of such a plateau is in qualitative agreement with both
simulations and experiment, and also with predictions from low-density
expansions.

It remains difficult to compare the mechanism causing a nontrivial
high-force plateau, $\Delta\zeta_\infty>0$, in our present MCT approach
with the one by Squires and Brady, \cite{Squires.2005} but the two descriptions share a number of apparent similarities.
In Ref.~\citenum{Squires.2005}, it arises from a singular boundary layer,
where advection and diffusion (otherwise negligible far from the probe
at high force) compete close to the particle; transport of fluid particles
out of the boundary layer requires motion parallel to the probe surface,
i.e., perpendicular to the force. In MCT, probe fluctuations transverse
to the direction of the applied force need to be included. In either case,
fast diffusive motion of the host fluid particles, characterized by the
short-time self-diffusivity \cite{Squires.2005} or $\Gamma_s$ in the
schematic model, dominates the high-force friction. Structural correlations,
which would be characterized by the much smaller long-time diffusivity
respectively the slow relaxation time of the correlators in
Figs.~\ref{fgr:phipar} and \ref{fgr:phiperp}, dominate the quiescent
friction.


%\balance


% Acknowledgments

We thank J.~F.~Brady for valuable discussions.
This work was supported by Deutsche Forschungsgemeinschaft, SFB TR6 project
A7, Helmholtz-Gemeinschaft (HGF Hochschul-Nachwuchsgruppe VH-NG~406),
and Junta de Andaluc\'\i{}a (P09-FQM-4938). We acknowledge partial funding
by the Zukunftskolleg der Universit\"at Konstanz and by the German
Excellence Initiative.
%, in the Konstanz Research Group ``Analysis and Numerics of
%Evolution Equations with Applications in the Science''.

\footnotesize{
\bibliography{rsc} %your .bib file
\bibliographystyle{rsc} %the RSC's .bst file
}

\end{document}